\documentstyle[aps]{revtex}
\begin{document}

\renewcommand{\textfraction}{0.10} \renewcommand{\topfraction}{1.0}
\renewcommand{\bottomfraction}{1.0} \flushbottom
\title{Scaling Considerations in Ground State Quantum Computation}
\author{Ari Mizel, M. W. Mitchell, and Marvin L. Cohen}
\address{Department of Physics, University of California at Berkeley,
Berkeley, CA 94720, USA, and Materials Sciences Division, Lawrence
Berkeley National Laboratory, Berkeley, CA 94720, USA.  }

\date{\today} 
\maketitle

\begin{abstract}
We study design challenges associated with realizing a ground state
quantum computer.  In such a computer, the energy gap between the
ground state and first excited state must be sufficiently large to
prevent disruptive excitations.  Here, an estimate is provided of this
gap as a function of computer size.  We then address the problem of
detecting the output of a ground state quantum computer.  It is shown
that the exponential detection difficulties that appear to be present
at first can be overcome in a straightforward manner by small design
changes.
\pacs{03.67.Lx}
\end{abstract}

\section{Introduction}

Recently, there has been intense interest among researchers in the
possibility of designing quantum computers \cite{Deutsch1998} that
calculate using the remarkable properties of quantum mechanics
\cite{Shor1994,Grover1996,Grover1997}.  Although the potential power
of quantum computation algorithms is enticing, there is great
difficulty associated with actually fabricating a quantum computer in
the laboratory.  A variety of schemes have been suggested
\cite{CiracZoller,Monroe,Turchette,Gershenfeld,Chuang,Loss,Burkhard,Shnirman,Makhlin,Averin,Nakamura,Mooij,Kane,Ioffe} and progress has been encouraging, but the feasibility of realizing a
useful quantum computer is still unclear.

In a recent paper, we proposed a novel ``ground state quantum
computation'' approach that could circumvent some of the main problems
with traditional quantum computer designs \cite{Mizel1999}.  This approach
replaces the progress of a usual, time-dependent quantum computation
with a single, time-independent state.  To see how this works, let us
suppose that a quantum computation algorithm consists of the evolution
of $M$ qubits through $N$ steps defined by $2^M \times 2^M$ unitary
matrices ${\cal U}_j$, $j = 1,\dots,N$.  To define the state of $M$
qubits at one step of the algorithm requires $2^M$ amplitudes.  To
describe the $M$ qubits at every step of the algorithm, from before it
begins until after it ends, requires $(N+1) \times 2^M$ amplitudes.
If we do not demand that the qubits evolve simultaneously from step to
step, allowing qubit \#1 be at step 2 while qubit \#3 is at step $6$,
then $(2(N+1))^M$ amplitudes are required to map out the development
of the qubits.  Let us suppose that we collect these $(2(N+1))^M$
amplitudes into a state $\left| \Psi \right>$ defined on a Hilbert
space of dimension $(2(N+1))^M$.  This state will contain all of the
information in a time-dependent quantum computation, but the state
itself will be completely time-independent.  Instead of developing
through time in accordance with an algorithm, the state will develop
through Hilbert space in accordance with the algorithm.

How can we explicitly describe this development through Hilbert space?
The projection of $\left| \Psi \right>$ onto some $2^M$ dimensional
subspace will contain the $2^M$ amplitudes necessary to describe the
state of the $M$ qubits when they are all at step 0 and have undergone
no unitary evolution.  Let us call this projection $P_{0}\left| \Psi
\right>$.  More generally, let us call $P_{j}\left| \Psi \right>$ the
projection onto the $2^M$ dimensional subspace that describes the
state of the $M$ qubits when they are at step $j$, $j=0,\dots,N$.
Suppose that we define an operator $A_{j,0}$ that carries the $2^M$
basis vectors of the subspace associated with $P_{0}$ into the $2^M$
basis vectors of the subspace associated with $P_{j}$.  Then, $\left|
\Psi \right>$ develops in accordance with a quantum computation
algorithm provided that
\begin{equation}
\label{development}
P_{j}\left| \Psi \right> = 
{\cal U}_{j}{\cal U}_{j-1}\ldots{\cal U}_{1}
A_{j,0}P_{0} \left| \Psi \right>
\end{equation}
for $j = 1,\dots,N$.

This formal notion makes it possible to propose a new ground state
approach to quantum computation.  In ground state quantum computation,
we do not make a register of qubits evolve in time by subjecting it to
a series of time-dependent Hamiltonians.  Instead, we perform
calculations by manufacturing a Hamiltonian $H$ whose ground state
develops according to the equation (\ref{development}).  This is
described precisely in reference \cite{Mizel1999}.  In that paper, an
appropriate Hamiltonian $H$ is found.  It is comprised of a sum of (i)
one-body terms denoted like $h_{a}^{k}(U_{a,k})$ where $U_{a,k}$
indicates the unitary evolution of qubit $a$ at algorithmic step $k$
of the calculation and (ii) two-body terms designated
$h^{j}_{a,b}({\rm CNOT})$ associated with a controlled not of qubit
$b$ by qubit $a$ at step $j$.  In \cite{Mizel1999}, a possible physical
realization of $H$ is suggested using states localized on quantum dots
to comprise the $(2(N+1))^M$ dimensional Hilbert space.

This ground state, time-independent approach has the attractive
characteristic that it avoids traditional decoherence problems
associated with time evolution.  However, it does have its own
challenging aspects that need to be addressed.  Many of these
challenges concern the scaling of a ground state quantum computer --
the feasibility of making such a computer larger and larger.  In this
paper, we address two of the most important considerations involved in
increasing the size of a ground state quantum computer.  First, we
study how the energy gap between the ground state and first excited
state of the Hamiltonian $H$ will depend on $N$ and $M$.  Clearly, if
a ground state quantum computer is to function properly the gap must
be large enough in energy that the computer will reliably remain in
its ground state.  Second, we investigate the problem of measuring the
outcome of a ground state quantum computation.  In its most naive
form, a ground state quantum computer would become very difficult to
probe as it grew in size.  We propose several means of avoiding this
difficulty.

\section{Gap}

If a ground state quantum computer is raised into an excited state,
its wavefunction can no longer be relied upon to satisfy equation
(\ref{development}), and the computer therefore does not compute
correctly.  To avoid such excitations, the computer must possess a
sufficiently large energy gap between ground state and first excited
state.  The gap must be significantly larger than the available
thermal energy $k_BT$, for example.

As a ground state quantum computer grows in size, its gap will
decrease, limiting the length of practical computations.  Here, we
study this limit, describing the size dependence of the gap of the
Hamiltonian described in \cite{Mizel1999}.  We argue that the gap
shrinks approximately like $1/(N+1)^2$ and prove the existence of a
lower bound that scales as $1/(N+1)^4$.

\subsection{Single Qubit}

To obtain these quantitative estimates of the gap, we first consider
the case of a single qubit computer.  Here, the Hamiltonian is simply
$H = \sum_{i=1}^{N} h^{i}(U_{i})$ where \cite{Mizel1999}
\begin{equation}
h^{i}(U) \equiv \epsilon
 \left[
 C^{\dagger}_{i-1}C_{i-1} + C^{\dagger}_{i}C_{i} -
 \left(C^{\dagger}_{i} U C_{i-1} + {\rm h.c.}\right)\right].
\end{equation}
is associated with the development of the single qubit from step $i-1$
to step $i$ of the calculation.  It is convenient to make a unitary
transformation from the $C_i$ operators to new operators
$(\Pi_{j=1}^{i} U_j) C_i$.  This changes the form of the Hamiltonian
to $H = \sum_{i=1}^N h^{i}(I)$ where every $U_i$ that appears in $H$
has been replaced by the 2 by 2 identity matrix $I$.  To determine the
eigenspectrum of this new $H$, we solve the determinantal equation
$\hbox{det} (H-E) = 0 $.  The determinant is evaluated by deriving and
solving a recursion relation on matrices of increasing size.  We find
that
\begin{equation}
\label{deteqn}
\hbox{det} (H-E) = \epsilon ^{2(N+1)} \frac{k-2 + 1/k}{k-1/k}(k^{2(N+1)} - \frac{1}{k^{2(N+1)}}) 
\end{equation}
where $k = (1 - \frac{E}{2\epsilon}) + \sqrt{(1-\frac{E}{2\epsilon})^2
-1}$.  Setting the determinant (\ref{deteqn}) to zero yields
eigenenergies $E = E_m \equiv 2\epsilon(1-cos(\pi m/2(N+1)))$, where $m$
is an integer between 0 and $2N+1$. The ground state has energy $E=E_0
= 0$, and the first excited state has energy
$E=E_1=2\epsilon(1-cos(\pi/2(N+1))) \rightarrow \epsilon \pi^2/(2(N+1))^2$ for
large N.  Thus, the gap decreases as $1/(N+1)^2$.  This is true for a
single qubit, and also for any number of non-interacting qubits.

\subsection{One CNOT Gate}

Of course, a useful quantum computer must have interactions among
qubits, so the behavior of the gap must be examined when interactions
are present.  To begin, we address the case of exactly two qubits
interacting through exactly one CNOT gate.  The full Hamiltonian
includes single qubit $h^{i}(U_i)$ terms and one $h^{j}(CNOT)$
interaction term at stage j.

We begin by examining the Hamiltonian with the $h^{j}(CNOT)$ term
omitted.  Without the $h^{j}(CNOT)$ term, the computer has two
disjoint regions for each qubit, one ``upstream'' of the omitted CNOT
gate consisting of stages 0 to $j-1$ and one ``downstream'' of the
omitted CNOT gate consisting of stages $j$ to $N$.  Since an electron
in one of the disjoint regions will possess the eigenspectrum of a
single, non-interacting qubit, the first excited state in such a
region will have an amount of energy of order $1/(N+1)^2$.  If we neglect
such ``high-energy'' states, only the (doubly degenerate) ground
states of the two regions make important contributions to the
electronic state of each qubit.  This means that each qubit has four
available states, leading to an effective 16 dimensional Hilbert space
for the two qubit system.

It is straightforward to diagonalize the interaction Hamiltonian
$h^{j}(CNOT)$ analytically in this $16 \times 16$ basis.  The result
is a (fourfold degenerate) ground state of the computer with zero
energy, an (eightfold degenerate) first excited state with energy
$\epsilon / (j)(N-j+1)$, and a (fourfold degenerate) second excited
state with energy $\epsilon / (N-j+1)^2 + \epsilon /(j)(N-j+1) +
\epsilon / (j)^2$.  So, the energy of the gap in the 16 dimensional
Hilbert space scales as $\epsilon /(j)(N-j+1) \sim 1/N^2$.

What relationship does the gap in this 16 dimensional Hilbert space
have to the exact gap of the system?  The (fourfold degenerate) ground
state in this 16 dimensional Hilbert space is, in fact, the exact
ground state in the whole Hilbert space.  Hence, the (eightfold
degenerate) first excited state in the 16 dimensional Hilbert space is
actually orthogonal to the exact ground state of the system.  It follows
that the quantity $\epsilon / (j)(N-j+1)$ represents a rigorous
variational upper bound to the exact first excited state energy of the
system.

The variational upper bound should provide a reasonable estimate of
the true value of the gap.  However, for our purposes we are perhaps
more interested in having a guaranteed lower bound to the gap.  Such a
lower bound would ensure that, when less than a specified amount of
energy is available, the computer will not experience a disruptive
excitation.  As it turns out, it is possible to show that the gap has
a rigorous lower bound of $\alpha/(N+1)^{4}$ for some real positive
$\alpha$.  The following is an argument by contradiction.

We know the ground state of the Hamiltonian $H$ has energy zero.
Suppose that $\left| \psi \right>$ is some state of the 2-particle
system that is orthogonal the ground state of $H$.  Assume that the
expectation value $\left< \psi \right| H \left| \psi \right>$
satisfies
\begin{equation}
\label{assume}
\left< \psi \right| H \left| \psi \right> < \alpha/(N+1)^{4} \equiv
E_{\hbox{\small lower}}.
\end{equation}
To evaluate the left hand side and draw a contradiction, we split the
Hamiltonian $H$ into $H_0$, which consists of only single body terms,
and $H_1 = h^{j}(CNOT)$, which consists of only interaction terms.
Both $H_0$ and $H_1$, it is straightforward to show, are positive
semi-definite.  Consider the form of $\left| \psi \right>$ in a basis
of eigenstates of $H_0$
\begin{equation}
\label{decomp}
\left| \psi \right> = \sum _{n,i} c_{n,i} \left| \phi _{n,i} \right>
\end{equation}
where the $i$ labels the degenerate eigenstates with the $n$th
eigenenergy.  Saying that $\left| \psi \right>$ is orthogonal to the
(fourfold degenerate) ground state of $H$ essentially means that
$\left| \psi \right>$ has no contribution from the four states $\left|
\phi _{n,i} \right>$ for which $H_0+H_1 \left| \phi _{n,i} \right> =
0$. Therefore, $\left| \psi \right>$ can consist of a superposition of
eigenstates of $H_0$ with eigenenergies greater than zero and the 12
zero energy eigenstates of $H_0$ that are orthogonal to the (fourfold
degenerate) ground state of the system.

Further reflection shows that $\left| \psi \right>$ cannot involve
exclusively the 12 zero energy eigenstates of $H_0$ because then
$\left< \psi \right| H \left| \psi \right>$ would go like $1/(N+1)^2$, as
we showed above, violating the assumption (\ref{assume}). The
state must therefore possess some contributions from excited
eigenstates of $H_0$.  These states have eigenenergies of at least
$\epsilon \pi^2/(2(N+1))^2$, as we saw in our single qubit analysis, so
assumption (\ref{assume}) limits the contribution from such states
to $\sum_{n>0,i} |c_{n,i}|^2 < E_{\hbox{\small lower}}/(\epsilon
\pi^2/(2(N+1))^2)$.  This limit exists even though $H_1$ is present since
$H_1$ is positive semi-definite and cannot decrease the expectation
value produced by $H_0$.  Hence we find that
\begin{eqnarray}
\left< \psi \right| H \left| \psi \right> & = & \sum _{i,j} c_{n=0,i}^*
c_{n=0,j} \left< \phi _{n=0,i} \right| H_0 + H_1 \left| \phi _{n=0,j}
\right> + \sum _{n>0,m>0,i,j} c_{n,i}^* c_{m,j} \left< \phi
_{n,i} \right| H_0 + H_1 \left| \phi _{m,j} \right> \\
& & + \sum _{m>0,i,j} \left( c_{m,i}^* c_{n=0,j} \left< \phi
_{m,i} \right| H_1 \left| \phi _{n=0,j} \right> + c_{n=0,i}^*
c_{m,j} \left< \phi _{n=0,i} \right| H_1 \left| \phi _{m,j}
\right> \right) \\
& > & \frac{\epsilon}{(N+1)^2}(1-\sum_{n>0,i} |c_{n,i}|^2) + \frac{\epsilon \pi^2}{(2(N+1))^2} \sum_{n>0,i} |c_{n,i}|^2 - 2\left|\sum_{i} c_{n=0,i}\right|\left|\sum_{m>0,i} c_{m,i}\right| \frac{\mu}{(N+1)^2} \\ 
& > & \frac{\epsilon}{(N+1)^2} + 0 - 2\left|\sqrt{12}\right| \left|\sqrt{4(N+1)^2-16} \sqrt{\frac{E_{\hbox{\small lower}}}{\epsilon
\pi^2/(2(N+1))^2}}\right| \frac{\mu}{(N+1)^2}
\end{eqnarray}
here $-\mu/(N+1)^2$ is the most negative value of $\left< \phi _{m>0,i}
\right| H_1 \left| \phi _{n=0,j} \right>$.  This last inequality
contradicts the assumption (\ref{assume}), however, since
\begin{eqnarray}
 \frac{\epsilon}{(N+1)^2}  + 0 - 2\left|\sqrt{12}\right| \left|\sqrt{4(N+1)^2-16}
\sqrt{\frac{E_{\hbox{\small lower}}}{\epsilon
\pi^2/(2(N+1))^2}}\right| \frac{\mu}{(N+1)^2} & = & \\ \nonumber
\frac{1}{(N+1)^2}(\epsilon - 2\left|\sqrt{12}\right| \left|\sqrt{\frac{4(N+1)^2-16}{(N+1)^2}}
\sqrt{\frac{\alpha}{\epsilon \pi^2/4}}\right| \mu) & > & E_{\hbox{\small lower}}
\end{eqnarray}
provided that $\alpha$ is chosen to be sufficiently small.  This
contradiction shows that the assumption (\ref{assume}) is not
valid.  Since the ground state has energy zero, and any state
orthogonal to the ground state has energy at least $E_{\hbox{\small
lower}}$, we have a lower bound $E_{\hbox{\small lower}}$ on the value
of the gap.

\subsection{Arbitrary Computer}

It is straightforward to apply these results to the case of $M$ qubits
interacting via an arbitrary number of CNOT gates.  First of all, the
variational upper bound on the gap of $\epsilon / (j)(N-j+1)$ still
holds.  This is because different CNOT Hamiltonians
$h^{j}_{a,b}(CNOT)$ commute with one another, so we can treat each
separately when diagonalizing in a basis of zero energy, $M$-particle
eigenstates of $H_0$.

The lower bound on the gap of order $1/(N+1)^4$ also still holds,
which we demonstrate in the following way.  The Hamiltonian consists
of $H_0$, that governs the single qubit development between CNOT gates
and the CNOT gate terms $h^{j}_{a,b}(CNOT)$ themselves.  We begin by
dividing $H_0$ into parts labelled $(H_0)^{j}_{a,b}$ where the index
${j,a,b}$ suggests proximity to the CNOT gate controlled by
Hamiltonian $h^{j}_{a,b}(CNOT)$.  Let $(H_0)^{j}_{a,b}$ consist of
terms that control the single qubit development of qubit $a$ between
CNOT gate ${j,a,b}$ and the previous CNOT experienced by qubit $a$,
terms that control the single qubit development of qubit $a$ between
CNOT gate ${j,a,b}$ and the next CNOT experienced by qubit $a$, terms
that control the single qubit development of qubit $b$ between CNOT
gate ${j,a,b}$ and the previous CNOT experienced by qubit $b$, and
terms that control the single qubit development of qubit $b$ between
CNOT gate ${j,a,b}$ and the next CNOT experienced by qubit $b$.  By
this definition of the $(H_0)^{j}_{a,b}$, we have $H_0 = \frac{1}{2}
\sum _{j,a,b} (H_0)^{j}_{a,b} + $ extra positive semi-definite one
body terms that are near the first or last stages of the computer.

Now, with this division described, we are in a position to demonstrate
the lower bound of order $1/(N+1)^4$.  Suppose that some $M$-particle
state $\left| \psi \right>$ is orthogonal to the ($2^M$ fold
degenerate) ground state of the system.  It is possible to write
$\left| \psi \right>$ in the form
\begin{equation}
\left| \psi \right> = \sum _{n,i} c_{n,i} \left| \phi _{n,i} \right>
\end{equation}
where the $\left| \phi _{n,i} \right>$ are $M$-particle eigenstates of
$H_0$.  Because $\left| \psi \right>$ is orthogonal to the ground
state of the system, each term $\left| \phi _{n,i} \right>$ that
appears in $\left| \psi \right>$ must satisfy $(H_0)^{j}_{a,b} +
h^{j}_{a,b}(CNOT) \left| \phi _{n,i} \right> \ne 0$ for some CNOT gate
${j,a,b}$.  Let us call $\left| \psi^{j}_{a,b} \right>$ the sum of the
components $ c_{n,i} \left| \phi _{n,i} \right>$ that satisfy
$(H_0)^{j}_{a,b} + h^{j}_{a,b}(CNOT) \left| \phi _{n,i} \right> \ne
0$.  If any $\left| \phi _{n,i} \right>$ could belong in more than one
$\left| \psi^{j}_{a,b} \right>$, we include it in every possible
$\left| \psi^{j}_{a,b} \right>$.  Then,
\begin{eqnarray}
\left< \psi \right| H \left| \psi \right> & \ge & \sum _{j,a,b} \left<
\frac{1}{2} \psi^{j}_{a,b} \right| (H_0)^{j}_{a,b} + h^{j}_{a,b}(CNOT) \left| \psi^{j}_{a,b} \right>
\\ & \ge & \frac{1}{2} \sum _{j,a,b} \left< \psi^{j}_{a,b}
\right|\left. \psi^{j}_{a,b} \right> \frac{\alpha}{(N+1)^4} \ge
\frac{\alpha}{2(N+1)^4} \sim \frac{1}{(N+1)^4},
\end{eqnarray}
where we have made use of the lower bound $\alpha/(N+1)^4$ derived in
the last section.  The last inequality holds because every component
of $\left| \psi \right>$ appears in at least one of the $\left|
\psi^{j}_{a,b} \right>$.  This result shows that a lower bound $\sim
1/(N+1)^4$ holds for an arbitrary number of qubits and CNOT gates.

\section{Detection}

The task of detecting the result of a ground state computation seems
daunting at first.  Each qubit in the computer must be measured in the
final stage, which it only visits with probability $1/(N+1)$.  Since there
are $M$ qubits, the probability of making a successful measurement
scales as $1/(N+1)^M$.  This problem is not, however, unsurmountable.  In
the following we describe several schemes that circumvent the
detection scaling problems.

In the case of Grover's algorithm, or any other algorithm for which
the final state of the qubits factors, it is actually possible to
guarantee a successful measurement.  This is accomplished by placing
an additional readout electron just after the final stage of each
qubit, as in Fig. \ref{readout}.  Since the qubit electron will be
localized in one of the two dots in the final stage, the readout
electron will tend to get pushed toward the other dot by the Coulomb
interaction.  In the ground state of the entire system, qubit
electrons and readout electrons included, a readout electron will be
localized to the left (right) if its qubit is in the right (left) at
the final stage.  The outcome of the algorithm can always be
determined by detecting the position of the readout electrons, even
though the qubit electrons have only a small probability of residing
at the final stage.  In effect, the readout electrons are providing a
continuous, passive measurement of the final state of the computer.
Although this readout electron method is guaranteed to work for an
algorithm for which the final state of the qubits factors, it will not
work at all if the final state of the qubits does not factor.  For a
general algorithm, another approach is necessary.

An approach which applies to arbitrary algorithms involves the
adjusting the Hamiltonian at the final stage $N$ for each qubit $a$.
Suppose that the operator $C^{\dagger}_{a,N}$ is replaced by $\beta
C^{\dagger}_{a,N}$ everywhere it appears in the Hamiltonian and
$C_{a,N}$ by $\beta C_{a,N}$, where $\beta$ is a small fraction.  All
algorithms will still work just as before, but we are ``tipping'' the
computer toward the final stage so that the qubits reside there more
often.  Then, it follows that in the ground state of the system, each
qubit has $1/\beta$ times greater amplitude on the final stage than on
the previous stages.  The probability of detecting all qubits on the
final stage is of order $1/(1+ \beta ^2 N)^M$.  If $\beta$ is set to
be of order, say $1/\sqrt{MN}$, we find that the probability of all
qubits being at the final stage goes as approximately $(1-1/M)^M$,
which approaches $\exp (-1)$.  It only takes two or three attempts to
catch all of the qubits at the final stage.

Of course, the change in the Hamiltonian will effect the gap.  If the
final operators are scaled by a factor $\beta$ then the quantity
$\hbox{det}(H-E)$ of non-interacting qubits will change to
\begin{equation}
\label{deteqn2}
\hbox{det} (H-E) = \epsilon ^{2(N+1)} \frac{k-2 + 1/k}{k-1/k}(k^{2(N+1)} -
\frac{1}{k^{2(N+1)}} + (\beta^2 - 1)(k^{2N+1} -  \frac{1}{k^{2N+1}}))
\end{equation}
Setting this determinant to zero, we find that the gap of
non-interacting qubits still scales roughly as $1/(N+1)^2$ for
arbitrary $\beta$ between zero and one.  Once CNOT gates are included,
however, the gap will have a variational upper bound of order
$1/(N+1)(N+1/\beta^2)$ and a lower bound of $E_{\hbox{\small lower}} =
\alpha/(N+1)^2(N+1/\beta^2)^2$.  If $\beta = 1/\sqrt{MN}$, then the upper
bound is $1/(NM+N)(N+1)$ and the lower bound $E_{\hbox{\small lower}} =
\alpha /(N+1)^2(NM+N)^2$.

Another technique for alleviating measurement problems is to
``synchronize'' the arrival of the qubits at the final stage.  In our
CNOT gate, the target electron cannot proceed beyond the gate until
the control electron has.  A controlled ``identity'' (CID) gate could
be constructed that would function similarly, preventing a target
electron from proceeding beyond the gate until after a control
electron, but always subjecting the target qubit to an identity
operation and never a NOT operation.  With this gate, the arrival of
the qubits at the final stage could be ``synchronized.''  Suppose that
each qubit controls the entry of the next qubit to a ground state
quantum computer's final stages, using a CID gate.  Then, whenever
qubit $M$ is found in the final stage, all qubits are there.  This
could be useful for detection schemes, although it would not enhance
the overall probability of finding the qubits at the final stage of
the computer.

\section{Conclusion}

In this paper, we have explored some important challenges to
constructing a ground state quantum computer.  First, we found upper
and lower bounds for the energy gap between the computer's ground
state and first excited state.  The bounds provide guidelines to
making a computer of a specified size that can be relied upon to
remain in its ground state.  Next, several schemes were presented for
easing qubit detection difficulties.  These schemes indicate how to
probe a ground state quantum computer so that it will yield output
with certainty or at least high probability.  It is hoped that the
analysis of energy gap and detection in this paper complements our
initial proposal and eases the task of designing and fabricating a
ground state quantum computer in the laboratory.

\acknowledgements

We gratefully thank D. A.  Lidar and V. J. Mizel for helpful comments
and references.

This work was supported by National Science Foundation grant
No. DMR-9520554, and the Director, Office of Energy Research, Office
of Basic Energy Services, Materials Sciences Division of the
U.S. Department of Energy under Contract No. DE-AC03-76SF00098; and
the Office of Naval Research grant No. N000149610034.

\begin{figure}
\leavevmode
\caption{Additional readout electrons are placed after the final
stage of each qubit.  The position of these additional electrons
indicates the outcome of the algorithm.}
\label{readout}
\end{figure}

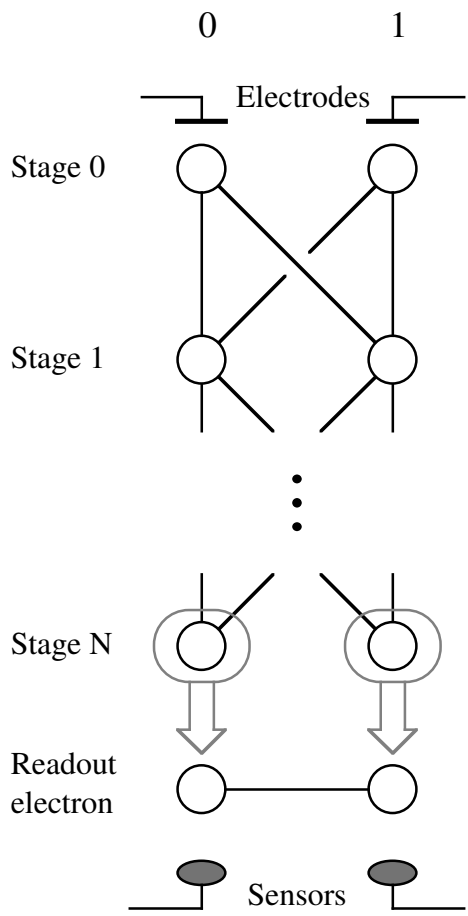
\end{document}